\documentclass{article}
\usepackage{amsmath}
\usepackage{amssymb}
\usepackage{amsthm}
\usepackage{graphicx}
\usepackage{psfrag}
\usepackage[hang, nooneline]{subfigure}

\author{Natascha Riahi \footnote{e-mail address: natascha.riahi@gmx.at}
\\ University of Vienna, Faculty of Physics, Gravitational Physics
\\ Boltzmanng. 5, 
1090 Vienna, Austria}
\date{}
\title{Wavepacket evolution in unimodular quantum cosmology }

\begin{document}

\maketitle

\begin{abstract}
The unimodular theory of gravity admits a canonical quantization of minisuperspace
models without the problem of time. We derive instead a kind of Schr\"odinger equation.
We have found unitarily evolving wave packet solutions for the special case of a
massless scalar field and a spatially flat Friedmann universe. We show that the longterm
behaviour of the expectation values of the canonical quantities corresponds to the
evolution of the classical variables. The solutions provided in an explicit example can
be continued beyond the singularity at t=0, passing a finite minimal extension of the
universe.
\end{abstract}

\section{Introduction}
The canonical quantization of general relativity leads to the so-called  problem of time (see \cite{Kiefer} and references therein).
In most non-perturbative approaches of quantum gravity time has disappeared from the theory and is seen as an artifact of the classical limit.
In contrast to this we will discuss here the quantization of a minisuperspace model in the framework of unimodular gravity. This theory is practically equivalent to general relativity at the classical level, but since it has a different canonical structure
time does not disappear from the quantum theory (\cite{HT}).
Investigations of unimodular quantum cosmology can be found in \cite{Unruh},\cite{DLS} as well as more recently in 
\cite{Chiou} where unimodular quantum loop
cosmology is discussed. 
In \cite{DLS} a semiclassical wave function  via path integral for an empty universe with positive curvature is 
constructed.
The time evolution fails to be unitary. The model mentioned in \cite {Unruh} is the spatially flat universe with a massless scalar field.
Here
the general solution is given only formally, and the properties of the wave packet evolution 
in particular  the question of unitary time evolution are not discussed. The solutions in  \cite{Chiou} apply for a flat universe filled with exotic matter with the equation of state $p=-2 \rho c^2$.

In this article we consider the quantization of spatially flat universe with a massless scalar field. We construct a class of unitarily evolving solutions  
with a negative expectation value of the Hamiltonian (that correspond classically to an infinitely expanding universe). 
Based on an example we investigate the time evolution of characteristic expectation values and compare it to the classical dynamics. 
\section{Unimodular theory}
The Einstein-Hilbert action of general relativity is given by

\begin{equation}
\label{EH}
S_{EH}=\frac{1}{2 \kappa}\int _{\mathcal{M}}d^4x\,\sqrt{-g}\,(R-2 \Lambda)-
\frac{1}{ \kappa} \int _{\mathcal{\partial M}}d^3x\,\sqrt{h}\,K\,,
\end{equation}
where 
\[
 \kappa=\frac{8 \pi G}{c^4}
 \]
contains the velocity of light $c$ and the gravitational constant $G$. The second integral is defined
on the spacelike boundary $\mathcal{\partial M}$ of the considered space-time region $\mathcal{M}$. The space-time metric
$g_{\mu \nu}$ with $det\, g_{\mu \nu }\equiv g$ induces a three-dimensional metric  $h_{a b}$ with $det\, h_{a b }\equiv h$
on the boundary $\mathcal{\partial M}$. The corresponding second fundamental form is denoted by $K_{a b}$ with the trace $K$.
If we also take into account  the matter action $S_{m}$ that describes the fields,
the variation of $S_{EH}+S_{m}$ with respect to the metric
yields the Einstein equations.

\begin{equation}
\label{GR}
 R_{\mu \nu}-\frac{1}{2}g_{\mu\nu}R=\kappa\, T_{\mu \nu}-\Lambda \,g_{\mu \nu },
\end{equation}

where the energy- momentum tensor is given by
\begin{equation}
\label{EM}
 T_{\mu \nu}=-
 \frac{2}{\sqrt{-g}}
 \frac{\delta S_{m}}{\delta g^{\mu \nu}}\,.
\end{equation}
If we start instead with an Einstein Hilbert action (\ref{EH}) with $\Lambda=0$ and vary it under the restriction  $-g=1$, we obtain Einsteins equations with an arbitrary additional
constant $\Lambda$ , that can be identified with the cosmological constant of general relativity (\cite{HT}). 
\begin{subequations}
\label{Unimod}
\begin{align}
& R_{\mu \nu}-\frac{1}{2}g_{\mu\nu}R=\kappa\, T_{\mu \nu}-\Lambda \,g_{\mu \nu } \label{Unimoda}\\
& \sqrt{-g}-1=0 \label{Det} \,.
\end{align}
\end{subequations}
This theory is called unimodular gravity.
Any solution of unimodular gravity (\ref{Unimod}) is also a solution of general relativity (\ref{GR})
for a specific cosmological constant and vice versa. The only difference between the two theories is, that $\Lambda$ is
a natural constant in general relativity while it is a conserved quantity in unimodular gravity. But since
in both theories the cosmological constant can not vary over the whole universe, we would have to investigate
different universes to determine if solutions with different $\Lambda$ exist (unimodular theory) or if
$\Lambda$ is a ''true'' natural constant.  So the two theories are practically indistinguishable.
Nevertheless the canonical structure of the theories differs (\cite{HT})and therefor the quantization of unimodular theory 
yields different results compared to the quantization of general relativity (\cite{Unruh}). In this article we will confine
the discussion of the canonical structure to the minisuperspace model we wish to quantize.
\section{The spatially flat Friedmann universe with a scalar field in unimodular theory}
The metric of a homogeneous and isotropic spacetime (Friedmann universe)

\begin{equation}
\label{MFriedmann}
  ds^2=-N^2(t) c^2 dt^2+a^2(t) d\Omega^2_{3} 
  \end{equation}
  is characterized by the lapse function N(t) and the scale factor $a(t)$. If the spatial curvature is zero,
$d\Omega^2_{3}$ is the line element of three-dimensional flat space.

Inserting the metric into the Einstein-Hilbert action (\ref{EH}) with $\Lambda=0$ yields (\cite{Kiefer})
\begin{align}
& S_{EH}=\frac{3}{ \kappa}\int dt\,N\, \left(-\frac{\dot{a}^2 a}{c^2 N^2}\right)\,v_{0}\,,
\nonumber\,
\end{align}
where $v_{0}$ is is the  volume of the spacelike slices according to (\ref{MFriedmann}).

The action of a scalar field in curved spacetime reads
\begin{equation}
\label{Smatter}
S_{m}=\int _{\mathcal{M}}d^4x\,\sqrt{-g}\,
\left(-\frac{1}{2} g^{\mu \nu} \phi_{,\mu}\phi_{,\nu}-V(\phi)\right)\,.
\end{equation}

If 
\[
V(\phi)=\left(\frac{m_{0} c}{\hbar}\right)^2 \frac{\phi^2}{2}
\]
the variation with respect to $\phi$ yields the Klein-Gordon equation
for a particle with mass $m_{0}$  
\begin{equation}
\label{KG}
\triangledown_{\mu}\triangledown^{\mu} \phi=\left(\frac{m_{0} c}{\hbar}\right)^2 \phi \,.
\end{equation}
Since we consider a spatial homogeneous spacetime, the
spatial derivatives of the field must be zero. Inserting (\ref{MFriedmann}), we find
\begin{equation}
\label{Sm}
S_{m}=\int\, dt\, N a^3 \left(\frac{\dot{\phi}^2}{2 N^2 c^2}-V(\phi) \right) v_{0}\,.
\end{equation}
Therefore the Lagrange function reads

\begin{equation}
\label{Lagrange}
L=v_{0}\epsilon N\,\left(-\frac{\dot{a}^2 a}{c^2 N^2}\right)+
v_{0} N a^3\left(\frac{\dot{\phi}^2}{2 N^2 c^2}-V(\phi)\right)\,,
\end{equation}

where we have introduced the abbreviation $\epsilon \equiv 3/( \kappa)$. If we incorporate $v_{0}$ into the variables $a$ and $N$
as well as $V(\phi)$ 
\begin{equation}
\label{Skalierung}
a \rightarrow v_{0}^{\frac{1}{6}} a\,,\quad
N \rightarrow v_{0}^{-\frac{1}{2}} N\,,\quad
V(\phi)\rightarrow v_{0} V(\phi)\,,\quad
\end{equation}
we find for the rescaled Lagrangian
\begin{equation}
\label{LagrangeR}
L=\epsilon N\,\left(-\frac{\dot{a}^2 a}{c^2 N^2}\right)+
N a^3\left(\frac{\dot{\phi}^2}{2 N^2 c^2}-V(\phi)\right)\,.
\end{equation}
According to unimodular theory the lapse function (\ref{Luni}) is determined by
$N=a^{-3}$ and the Friedmann metric (with unscaled quantities) has the form

\begin{equation}
\label{MetrikU}
ds^2=-(c^2/ a^6(t)) dt^2+a^2(t)d\Omega^2_{3}
\end{equation}

Since the condition $N a^{3}=1$ is not influenced by the scaling
(\ref{Skalierung}) we find for the rescaled unimodular
 Lagrange function (\ref{LagrangeR}) 

\begin{equation} 
\label{Luni}
L_{uni}=\epsilon (-\frac{\dot{a}^2 a^4}{c^2})
+a^6\frac{\dot{\phi}^2} {2 c^2}-V(\phi)\,.
\end{equation}

The momenta conjugate to the variables $a$ and $\phi$ read
\begin{equation}
\label{Impulse}
 p_{\phi}=\frac{1 }{c^2}\dot{\phi} a^6\qquad p_{a}=-\frac{2  \epsilon}{c^2}\dot{a}a^4 \,.
\end{equation}

We obtain for the Hamiltonian of the unimodular theory

\begin{equation}
\label{Huni}
 H_{uni}=\frac{c^2}{2}\frac{p_{\phi}^2}{a^6}-\frac{c^2}{4 \epsilon}\frac{p_{a}^2}{a^4}
 + V(\phi)
 \end{equation}
The Hamiltonian is a conserved quantity. If we write the Hamiltonian
as a function of the configuration variables $a, \phi$ and their derivatives
\begin{subequations}
\label{Constraint}
\begin{equation}
H_{uni}=\frac{1}{2 c^2}\dot{\phi}^2 a^6-\frac{\epsilon}{c^2}\dot{a}^2 a^4+V(\phi)
\end{equation}
we find that 
\begin{equation}
H_{uni}\equiv -\frac{\Lambda \epsilon}{3}
\end{equation}
\end{subequations}
equals the Hamiltonian constraint of general relativity for
$N=1/a^3$. We see that this special labeling of the conserved quantity makes the solutions
of unimodular gravity and general relativity coincide (see Appendix \ref {Hamiltonian}). Nevertheless the Hamiltonian according to general relativity differs from (\ref{Huni}).
\section{Classical solutions of a flat Friedmann universe with a massless scalar field}
The simplest case of a matter Lagrangian is
$V=0$, which would correspond to the field of a massless particle with spin zero (\ref{KG}). If instead a perfect fluid matter model is 
chosen, 
the solutions for a massless scalar field can be shown to be equivalent to the solutions for the special case of 
stiff matter (see \cite{Masden}).

The Hamiltonian reads
\begin{equation}
\label{H0}
 H_{uni}=\frac{c^2}{2}\frac{p_{\phi}^2}{a^6}-\frac{c^2}{4 \epsilon}\frac{p_{a}^2}{a^4}
\,.
\end{equation}
According to the equations of motion $p_{\phi}$ is a conserved quantity 
\begin{equation}
\label{Erhaltung}
\dot{p}_{\phi}=0
\end{equation}
and the time-dependence of the field is given by
\begin{equation}
\label{k}
 \dot{\phi}=\frac{p_{\phi} c^2}{a^6}\,.
\end{equation}

If we assume $p_{\phi}=0$, we obtain the de-Sitter solutions  with $\Lambda\,>\,0$. The conservation of the Hamiltonian
\[
 -\frac{c^2}{4 \epsilon}\frac{p_{a}^2}{a^4}=H_{uni}=-\frac{\Lambda \epsilon}{3}
\]

implies
\[
 \frac{\epsilon}{c^2}(\dot{a} a^2)^2=\frac{\Lambda \epsilon}{3}\,.
\]
We then find for the scale factor
\begin{equation}
 \label{deSitterb}
 a=(\sqrt{3 \Lambda} c t)^{\frac{1}{3}}\,.
\end{equation}

Note that this solution of unimodular theory coincides with a solution of
general relativity with a choice of coordinates 
with $N=1/a^3$\, in (\ref{MFriedmann}).

If $p_{\phi} \neq 0$, 
\[
 -\frac{c^2}{4 \epsilon}\frac{p_{a}^2}{a^4}+\frac{c^2 p^2_{\phi}}{2 a^6}=-\frac{\Lambda \epsilon}{3}
\]
yields

\[
 \frac{\dot{a}^2 a^{10}}{c^2}=\frac{\Lambda}{3}a^6+\frac{1}{\epsilon}\frac{p_{\phi}^2 c^2}{2 }\,.
\]
We obtain
\begin{subequations}
 \begin{align}
& a(t)=\left(6\sqrt{\frac{ p_{\phi}^2}{2 \epsilon}} c t+3 \Lambda\, c^2\,t^2 \right)^{1/6}
 \label{Skala}\\
 & \phi(t)=\frac{1}{6}\sqrt{2 \epsilon} Sign[p_{\phi}]
\mbox{ln} 
\left[
 \frac{t \,Z}{1+\frac{\Lambda}{2}
 \sqrt{\frac{2\, \epsilon}{ p_{\phi}^2}}\,
 t} 
 \right]\,,
 \end{align}
  \end{subequations}
where Z is an integration constant that determines $\phi(0)$. For the scale factor we have assumed $a(0)=0$.

\section{Quantization of a flat Friedmann universe with a massless scalar field}
The canonical quantization of the unimodular Hamiltonian of the model
(\ref{H0}) yields for the massless case ($V=0$)

\begin{equation}
 \label{QImplulse}
 \hat{p}_{a}=-i \hbar \frac{\partial}{\partial a}\,,\quad
  \hat{p}_{\phi}=-i \hbar \frac{\partial}{\partial \phi}\,,
\end{equation}

\[
\widehat{H}=\frac{\hbar^2 c^2}{4 \epsilon}\frac{1}{a^5}\frac{\partial}{\partial a} a \frac{\partial}{\partial a}
 -\frac{\hbar^2 c^2}{2} \frac{1}{a^6 }\frac{\partial^2}{\partial \phi^2}\,.
\]

Here we have chosen the factor ordering that gives the part of the Hamiltonian that is quadratic in the momenta the form of a 
Laplace Beltrami 
operator (\cite{Kiefer})

The evolution of the wavefunction  $\psi(a,\phi,t)$ is  determined by
\begin{equation}
\label{S}
\widehat{H}\psi=i \hbar \frac{\partial}{\partial t} \psi\,.
 \end{equation}
 The Hamiltonian is symmetric with respect to the inner product defined by the measure $a^5 da d\phi$, where 
 $a \in (0,\infty)$ and $\phi \in (-\infty,\infty)$.

With the transformation
\begin{equation}
\label{Trafo1}
 A=a^3/3 \qquad B= \frac{3}{\sqrt{2 \epsilon}}\phi\,,
\end{equation}
and the volume element   $3 \sqrt{2 \epsilon}A dA dB$,
the Hamiltonian assumes the form
\[
 \widehat{H}=\frac{\hbar^2 c^2}{4 \epsilon}
\left\{ \frac{1}{A}\frac{\partial}{\partial A}A\frac{\partial}{\partial A}-
 \frac{1}{A^2}\frac{\partial^2}{\partial B^2}\right\}\, .
\]

This expression has the appearance of the wave equation in polar coordinates, which  shows that our minisuperspace
is flat. The Hamilton operator equals the wave operator in Rindler spacetime. 
We  bring the Hamiltonian  into the simplest form
using a transformation to light-cone coordinates:
\begin{equation}
 \label{Trafo2}
u=A e^{-B}\qquad v=A e^{B}\,.
\end{equation}
We obtain the Hamiltonian
\begin{equation}
\label{Hop}
 \widehat{H}=\frac{\hbar^2 c^2}{\epsilon}\frac{\partial^2}{\partial u \partial v }
\end{equation}

The volume element is given by
 $\frac{\sqrt{\epsilon}}{2}du dv$ and $u \in (0,\infty)$,
$v \in (0,\infty)$. This is equivalent to a volume element $du\, dv$, if the wave functions are accordingly normalized.
We will search for solutions of ( \ref{S}) that are square integrable. Moreover they should fulfill the condition for a 
unitary time evolution that is given by
 \begin{equation}
\label{sa}
\langle \psi(t_{1})|\,\widehat{H} |\,\psi(t_{2}) \rangle\,=
\,\langle \widehat{H}\,\psi(t_{1})\,|\psi(t_{2}) \rangle\quad \forall\,t_{1},t_{2}\,.
\end{equation}
For $t_{1}=t_{2}$ this condition ensures that the norm of the 
 wavepackets is preserved:
\begin{equation}
 \frac{d}{dt}\langle\psi|\psi\rangle=0\,.
\end{equation}
Differentiating (\ref{sa}) with respect to $t_{2}$ 
results in the conservation of the expectation value of the Hamiltonian
\[
 \frac{d}{dt}\langle\psi|\widehat{H}|\psi\rangle=0\,.
\]
Higher time derivatives yield
\[
 \frac{d}{dt}\langle\psi|\widehat{H}^n|\psi\rangle=0\quad \mbox{for}
\quad n=2,3,\ldots\,.
\]

For the Hamiltonian (\ref{Hop}) the condition  (\ref{sa}) is equivalent to
\begin{equation}
\label{Bedingung}
\int_{0}^{\infty}\psi^{*}(0,v,t_{1})\frac{\partial}{\partial v}\psi(0,v,t_{2})dv-
\int_{0}^{\infty}\psi(u,0,t_{2})\frac{\partial}{\partial u}\psi^{*}(u,0,t_{1})du=0\,.
\end{equation}
If we choose  two real functions $f_{1}(x), f_{2}(x)$
where $f_{1}(0)= \pm f_{2}(0)$ then any solution of (\ref{S}) that obeys

\begin{equation}
\label{Rand}
 \psi(0,v,t)=C(t)f_{1}(v) \qquad
  \psi(u,0,t)=C(t)f_{2}(u)\,,
\end{equation}

 fulfills
(\ref{sa}). All solutions that evolve according to (\ref{Rand}) with an arbitrary function $C(t)$ and are square integrable at $t=0$ remain within
the linear subspace of the space of square integrable function characterized by (\ref{Rand}).
\section{Solutions with a negative (expectation) value of the Hamiltonian}
\label{Solutions}
Introducing
 \begin{equation}
 \label{tau}
 \tau=t \hbar c^2/\epsilon
 \end{equation}
we can write for the  Schr\"odinger equation (\ref{S})
\begin{equation}
\label{Sf}
\frac{\partial^2}{\partial u \partial v }\psi=i  \frac{\partial}{\partial \tau} \psi
\end{equation}
This equation looks rather simple and there are many possibilities to solve it. The challenge here is to find solutions that ensure a unitary time evolution. We will obtain this goal by a superposition of eigensolutions.
An  eigenstate $ \psi_{\Lambda}(u, v)$ with negative eigenvalue $h\equiv-\Lambda\,\epsilon/3<0$ fulfills the equation
\begin{equation}
\label{eigen}
\frac{\partial^2}{\partial u \partial v }\psi_{\Lambda}(u, v)=-\frac{\Lambda\epsilon}{3}\psi_{\Lambda}(u, v)\,,
\end{equation}
which gives the time-dependent solution
\begin{equation}
\label{eigensolution}
\psi(u,v,\tau)=\psi_{\Lambda}(u, v)e^{\frac{i \Lambda\,\epsilon \tau}{3}}\,.
\end{equation}

The Laplace transformation (see f.i. \cite{Za}) of (\ref{eigen}) with respect to $v$ 
yields

\begin{equation}
\label{Leigen}
s \frac{\partial L_{\Lambda}(u,s)}{\partial u}-
\frac{\partial \psi_{\Lambda}(u,0)}{\partial u} \,
= -\frac{\Lambda\epsilon}{3}L_{\Lambda}(u, 0)
\end{equation}
 
where  $L_{\Lambda}(u, s)$ is given by

\begin{equation}
L_{\Lambda}(u, s) = \mathcal{L}(\psi_{\Lambda}(u,\cdot)) = \int\limits_{0}^{\infty} \psi_{\Lambda}(u,v)  e^{-s v} dv \,,
\end{equation}
and we have used the differential rule of the Laplace transform (\cite{Za})
Solving (\ref{Leigen}), we obtain
\[
L_{\Lambda}(u,s)=
e^{-\frac{\Lambda \epsilon }{3 s}u}\,L_{\Lambda}(0,s)
+\int\limits_{0}^{u}
\frac{1}{s}
e^{-\frac{\Lambda \epsilon }{3 s}(u-x)}\,\frac{\partial \psi_{\Lambda}(u,0)}{\partial u}dx
\]
Applying the formula for the inverse Laplace transform (see \cite{Erdely})
\[
\mathcal{L}^{-1}\left\{\frac{1}{s}e^{-\frac{a}{s}}\right\}=
J_{0}\left(2 \sqrt{a x}\right)
\]
and the convolution theorem (\cite{Za}) we find for the inverse
Laplace transform
\begin{align*}
\psi_{\Lambda}(u,v)&=
\int_{0}^{v}
J_{0}\left[2\sqrt{\frac{\Lambda\,\epsilon u}{3}}\sqrt{v-x}\right]\,
\frac{\partial f_{1}}{\partial x}dx \\
&+\int_{0}^{u}
J_{0}\left[2\sqrt{\frac{\Lambda\,\epsilon v}{3}}\sqrt{u-x}\right]\,
\frac{\partial f_{2}}{\partial x}dx \\
&+J_{0}\left[2\sqrt{\frac{\Lambda\,\epsilon }{3}v u}\right]\,
f_{1}(0)\,,\\
&\mbox{where}\quad f_{1}(0)=f_{2}(0)\,.
\end{align*}
The functions $f_1(x),f_2(x)$ determine $\psi_{\Lambda}(u,v)$ at
the edges
\[
f_{1}(x)\equiv \psi_{\Lambda}(0,x) \qquad f_{2}(x)\equiv \psi_{\Lambda}(x,0).
\]
The superposition of the corresponding eigensolutions (\ref{eigensolution}) yields more general time-dependent wavepacket solutions of (\ref{Sf})
\begin{equation}
\label{Lambdap1}
\psi(u,v,\tau)=
\int_{0}^{\infty}e^{i \tau\,\epsilon \frac{\Lambda}{3}}
\psi_{\Lambda}(u,v)F\left(\Lambda\right) d\Lambda\,.
\end{equation}
The solution is then characterized by the functions $f_{1}(x),f_{2}(x),F(\Lambda)$. We assume that they can be chosen appropriately to ensure that $\psi(u,v,0)$ is square integrable.
We obtain for the time evolution at the edges
\begin{align}
\label{RandS}
 \psi(0,v,\tau)=C(\tau)f_{1}(v) \qquad
  \psi(u,0,\tau)=C(\tau)f_{2}(u) \\
  \mbox{where}\qquad C(\tau)=\int_{0}^{\infty}
  e^{i \epsilon\, \tau \frac{\Lambda}{3}}F(\Lambda)\,d\Lambda \,,
  \nonumber
\end{align}
which means that (\ref{Lambdap1}) meets the condition  (\ref{Rand}) for a unitary time evolution if $f_{1}(x),f_{2}(x)$ are real functions. Moreover it follows from the Riemann-Lebesgue lemma (see for instance \cite{Za})for the Fourier transform that 
\begin{equation}
\label{LimesC}
\lim_{\tau \rightarrow \infty}C(\tau)=0\,,
\end{equation}

if $F(\Lambda)$ is an absolutely integrable function. This implies for the wavefunction at the edges
\begin{equation}
\label{Limes}
\lim_{\tau \rightarrow \infty}
\psi(u,0,\tau)=
\lim_{\tau \rightarrow \infty}
\psi(0,v,\tau)=0 \,.
\end{equation}

It will turn out to be more convenient to represent $F(\Lambda)$
as 
\begin{equation}
\label{Hankel}
F(\Lambda)=\int_{0}^{\infty}
J_{0}\left[2 \sqrt{\frac{\epsilon\,\Lambda r}{3}}\right]
\frac{G(r)\,\epsilon}{3}\,dr\,.
\end{equation}

The initial wavefunction then reads
\begin{align}
\label{start}
\psi(u,v,0)&=
\int_{0}^{v}
G\left[u(v-x)\right]\,
\frac{\partial f_{1}}{\partial x}dx 
+\int_{0}^{u}
G\left[v (u-x)\right]\,
\frac{\partial f_{2}}{\partial x}dx 
+G\left[u v\right]\,
f_{1}(0)\,.
\end{align}

This is a consequence of the selfreproducing property of the Hankel transformation
(\cite{Za})
\begin{align}
\label{kernel}
&H(x)=\int_{0}^{\infty}\sqrt{x y}J_{0}( x y)
\int_{0}^{\infty}\sqrt{y z}J_{0}(y z)H(z)dz dy\,,
\end{align}
which is equivalent to
\begin{align*}
&G(\alpha)=\int_{0}^{\infty}J_{0}(2\sqrt{ \alpha \beta} ) \int_{0}^{\infty}J_{0}(\sqrt{\beta \gamma})G(\gamma)d\gamma d\beta\,,
&\quad \mbox{where}\quad G(\gamma)=H(\sqrt{2 \gamma})/\gamma^{\frac{1}{4}}\,.
\end{align*}
The relation (\ref{kernel}) applies to any absolutely integrable function $H(z)$ on $\mathbb{R}_{+}$ of bounded variation.
We also deduce from (\ref{Hankel},\ref{kernel}) that we can choose the function $G(z)$ in (\ref{start}) 
arbitrarily under the restriction that 
$G(z^2/2)\sqrt{z}$ is an absolutely integrable function of bounded variation.

Inserting (\ref{Hankel}) into (\ref{Lambdap1}) yields for the time-evolution

\begin{align}
\label{Lambdap2}
&\psi(u,v,\tau)= \\
&\frac{i}{\tau} \int_{0}^{\infty}\int_{0}^{u}
e^{-\frac{i v (u-x)}{\tau}-\frac{i r}{\tau}}
J_{0}\left[2 \frac{\sqrt{v (u-x)r}}{\tau}\right]
\frac{\partial f_{1}}{\partial x} dx G(r) dr 
\nonumber
 \\
&\frac{i}{\tau} \int_{0}^{\infty}\int_{0}^{v}
e^{-\frac{i u (v-x)}{\tau}-\frac{i r}{\tau}}
J_{0}\left[2 \frac{\sqrt{u (v-x)r}}{\tau}\right]
\frac{\partial f_{2}}{\partial x} dx G(r) dr
\nonumber 
\\
&\frac{i}{\tau} \int_{0}^{\infty}
e^{-\frac{i u v}{\tau}-\frac{i r}{\tau}}
J_{0}\left[2 \frac{\sqrt{u\, v\,r}}{\tau}\right]f_{1}(0)
  G(r) dr\,.
\nonumber
\end{align}

We have obtained this result applying the formula for the Laplace transformation
of a product of Bessel functions (\cite{Erdely})
\[
\int_{0}^{\infty}e^{-z s}J_{0}(2\sqrt{a z})J_{0}(2\sqrt{b z})dz=
\frac{1}{s}e^{-\frac{a+b}{s}}I_{0}\left(2 \sqrt{a b}\right)
\]
with the parameters
\[
s=-i\,\tau\,,\quad z=\epsilon \Lambda/3\,,\quad
a=r\,,\quad b=u(v-x)\,\quad
\mbox{or}\,\, b=v(u-x)\,\quad\mbox{or}\,\, b=v\,u\,\,.
\]
$I_{0}(x)$ is the modified Bessel function of the first kind (see for instance \cite{Magnus})and fulfills $I_{0}(i x)=J_{0}(x)$.
\section{Example for a wavepacket evolution with a negative expectation value of the Hamiltonian}

In section (\ref{Solutions}) we constructed solutions of (\ref{Sf}) with a  negative expectation value of the Hamiltonian and a
unitary time evolution. We had to require that the functions $G(z),f_{1}(x),f_{2}(x)$ ensure that the initial wave function 
(\ref{start}) is square integrable. Investigating an explicit example shows that an appropriate choice of these functions is possible. 
 The lengthy algebraic calculations
of this section were performed with Wolfram Mathematica.

We define

\begin{align}
& G(z)=e^{-z}\left(z^3-6 z^2+3 z+3\right) \\
 & f_{1}(x)=e^{-x}\left(-x^3+3 x^2\right)\\
 & f_{2}(x)=0
\end{align}

With the additional real parameters $\lambda $ and $\mu$, we find for the initial function according to (\ref{start})

 \begin{align}
 \label{Beispiel}
\psi(u,v,0)=\frac{1}{\sqrt{n_{0}}}
\int_{0}^{v}
G\left [\lambda u(v-x)\right]\,
\frac{\partial f_{1}(\mu x)}{\partial x}dx\,, 
\end{align}
where we have introduced a normalization factor that turns out to be
\[
 n_{0}=\frac{135}{3 \lambda}\,.
\]

We obtain for the expectation value of the Hamiltonian
\begin{equation}
 \langle \widehat{H} \rangle=-\frac{3 \lambda}{2}\cdot \frac{\hbar^2 c^2}{\epsilon}\,.
\end{equation}
According to (\ref{Trafo1},\ref{Trafo2}) the observable $A=u v$ is related to the 
scale factor by
\[
 A=u v=\frac{a^6}{9}\,.
\]

\begin{figure}[htp]
\centering
 \includegraphics[width=0.7\textwidth]{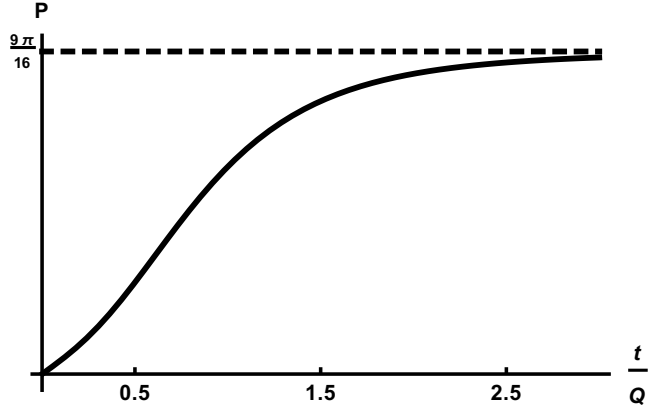} 
\caption{
The quantity  
$M \equiv \langle p_{\phi}\rangle  
\left(\frac{3 \hbar}{\sqrt{2 \epsilon}}\right)^{-1}$ 
converges to $\frac{9 \pi}{16}$.  We have chosen $\lambda=1\cdot m^{-3}$, $\mu=1\cdot m^{-3/2}$ and time is scaled by
$Q=\sqrt{\frac{2}{3 }} \cdot
\frac{\epsilon}{\hbar c^2 \lambda} $
}
\label{PP}
\end{figure}

The time evolution of $\langle A^2 \rangle$ is determined by (\ref{Lambdap2}):
\begin{align}
\label{a2}
&\langle A^2 \rangle\,= \frac{3}{40 \lambda(1 + \tau^2 \lambda^2)^2}\\
& \qquad \cdot\, \left[30 + 71 \tau^2 \lambda^2 + 67 \tau^4 \lambda^4 + 
     20 \tau^6 \lambda^6 + 
     15 \tau \lambda (1 + \tau^2 \lambda^2)^2 \mbox{arctan}(
       \tau \lambda)\right]\,. \nonumber
     \end{align}
We find for the late phase of the time evolution (where we have replaced $\tau$ by t (\ref{tau})):
\[
 \lim_{t \rightarrow \infty} \frac{d^2}{dt^2}  \langle A^2 \rangle\,=\,\frac{3 \lambda \left(\hbar c^2\right)^2}{\epsilon^2}
 =\,-\frac{2c^2}{\epsilon}\langle \widehat{H}\rangle\,,
\]

which  coincides with the classical behaviour according to (\ref{Skala}).
 We also find that the expectation value of $\widehat{p}_{\phi}$ approaches a constant value 
 \begin{equation}
 \label{Phi}
   \lim_{t \rightarrow \infty} \langle \widehat{p}_{\phi} \rangle\,=\,
   \frac{9 \pi}{16}\cdot \frac{3 \hbar}{\sqrt{2 \epsilon}}\,,
 \end{equation}
which shows that the classically conserved quantity $p_{\Phi}$ is only a constant of motion in the late phase of time evolution 
(see figure \ref{PP}).

\begin{figure}[htp]
\centering
 \includegraphics[width=0.7\textwidth]{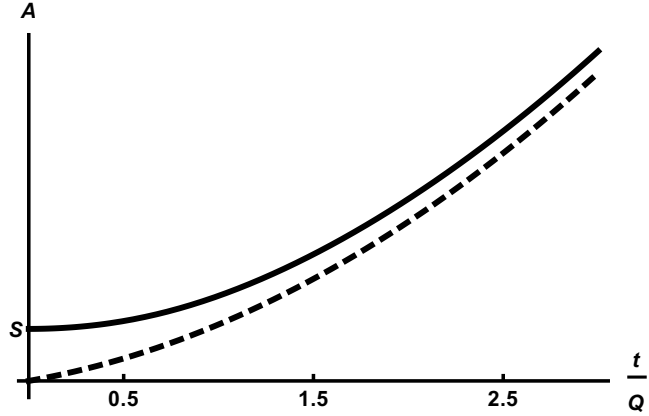} 
\caption{The time-evolution
of the expectation value of $A^2$ according to the initial wave packet (\ref{Beispiel}), compared to the classical evolution (dashed line). S is the minimal extension according to
the quantum evolution. $\mu\,,\lambda $ and $Q$ are defined as in figure \ref{PP}.
}
\label{Scalefactor}
\end{figure}

Nevertheless at the beginning of the time evolution, the observable $A^2$ that is proportional to
the sixth power of the scale factor assumes a constant value (\ref{Scalefactor}).
\[
S=\frac{9}{4 \lambda}\,.
\]

We can also prove that the linear part of the classical time evolution of $A^2$ (\ref{Skala})is 
reproduced by the
late time behaviour of the expectation value. Inserting for (\ref{Phi}), we find
\[
 \lim_{t \rightarrow \infty}\left(\frac{d}{dt}\langle A^2 \rangle
 -t \cdot\frac{d^2}{dt^2}\langle A^2 \rangle \right)=\frac{9 \pi}{16} \cdot \frac{\hbar c^2}{\epsilon}=
 p_{\phi}\cdot \frac{\sqrt{2}}{\sqrt{\epsilon} 3 \hbar}\,,
\]
which shows that the  time evolution of the expectation value approaches the classical evolution (see figure \ref{Scalefactor}).
For the variance of $A^2$ we find
\begin{align}
 &(\Delta A^2)^2\equiv\left\langle A^4 \right \rangle-\left\langle A^2 \right \rangle^2=\\
&\frac{768 + 
   t \lambda (27 \pi (9 + 14 t^2 \lambda^2 + 5 t^4 \lambda^4) + 
      2 t \lambda (1133 + 823 t^2 \lambda^2 + 
         128 t^4 \lambda^4)) - 
 54 t \lambda (9 + 14 t^2 \lambda^2 + 5 t^4 \lambda^4) ArcCot[
     t \lambda]}{(80( \lambda^2 + t^2 \lambda^4)}\\
   & \qquad  - \frac{
 9  (30 + 71 t^2 \lambda^2 + 67 t^4 \lambda^4 + 
    20 t^6 \lambda^6 + 
    15 t \lambda (1 + t^2 \lambda^2)^2 ArcTan[t \lambda])^2)}{
 1600 \lambda^2(1+ t^2 \lambda^2)^4}\,,
\end{align}

which yields
\begin{equation}
\lim_{t \rightarrow \infty} \frac{(\Delta A^2)^2}{t^4}=\frac{19 \lambda^2}{20}
\end{equation}

\section{Discussion and conclusions}
It is important to note that the system behaves more and more classically at late times. An approximate
classical behaviour far away from the singularity was also predicted for the case of exotic matter in
\cite{Chiou}. This is in contrast to
quantum mechanics, where
systems start with  classical motion developing quantum properties in the late phase of time evolution, as the revival phenomena 
show impressively (see f.i. \cite{Ro}). 
However this classical behaviour for late times  does not apply to all aspects of the time evolution 
because the uncertainty of the observable
$A^2$, that is related to the scale factor increases with time. We think that this would not happen with a more realistic matter model.

The construction of solutions according to (\ref{Lambdap1}) also works for negative times and so the solutions
can be continued beyond the classical singularity at $t=0$. In the special case we investigated (\ref{a2}) this 
would yield a
contracting universe that passes a minimal extension before  expanding again.
The question of solutions with a positive expectation value of the Hamiltonian remains open.
Moreover we think that unimodular quantum gravity gives the opportunity to investigate effects of cosmological evolution
(as inflation, acceleration of the expansion of the universe) in the framework of a quantum theory of gravity without being
confined to a semiclassical limit.
\section*{Acknowledgments}

I thank Helmut Rumpf for calling my attention to unimodular gravity.
I thank Christiane Lechner for proposing the transformation to light cone coordinates that helped
me to analyze the minisuperspace equation of unimodular quantum cosmology.

\appendix

\section{Hamiltonian constraint according to general relativity}
\label{Hamiltonian}
 Inserting the metric (\ref{MFriedmann}) into the Einstein-Hilbert action with cosmological constant yields
 \begin{align}
& S_{EH}=\frac{3}{ \kappa}\int dt\,N\, \left(-\frac{\dot{a}^2 a}{c^2 N^2}-\frac{ \Lambda}{3} a^3\right)\,v_{0}\,,
\nonumber\,.
\end{align}

Rescaling also $\Lambda$ 
\begin{equation}
\label{SkalierungL}
\Lambda \rightarrow v_{0}  \Lambda\,,\quad
\end{equation}

in addition to \ref{Skalierung}
we find for the rescaled Lagrangian
\begin{equation}
\label{LagrangeRG}
L=\epsilon N\,\left(-\frac{\dot{a}^2 a}{c^2 N^2}-\Lambda \frac{a^3}{3}\right)+
N a^3\left(\frac{\dot{\phi}^2}{2 N^2 c^2}-V(\phi)\right)\,.
\end{equation}

The variation of (\ref{LagrangeRG}) with respect to N yields 
\[
-\left(\frac{\dot{a}}{a N c}\right)^2 + \frac{\Lambda}{3}+\frac{1}{\epsilon}
\left(\frac{\dot{\phi}}{2 N^2 c^2}+V(\phi)\right)=0\,.
\]
This equation establishes a relation of the variables and their first derivatives. It represents the Hamiltonian constraint of 
general relativity and equals (\ref{Constraint}) for $N=1/a^3$. Nevertheless the Hamiltonian differs from the respective expression
in the canonical quantization of general relativity with cosmological constant (see also \cite{CS})
\begin{equation}
 \widehat{H}=-\frac{p_{a}^2 c^2}{4 a \epsilon}+\frac{p^2_{\phi} c^2}{2 a^3}+\frac{\Lambda a^3}{3}+a^3 V\left(\phi\right)\,.
\end{equation}

\end{document}